\documentclass[prb,twocolumn]{revtex4}

\usepackage{graphicx}
\usepackage{dcolumn}
\usepackage{amsmath}
\usepackage{epsfig}

\begin{document}

\title{Semiconductive and Photoconductive Properties of the Single Molecule Magnets
Mn$_{12}$-Acetate and Fe$_8$Br$_8$}

\author{J. M. North, D. Zipse, and N. S. Dalal}

\affiliation{Department of Chemistry and Biochemistry, and
National High Magnetic Field Laboratory, Florida State University,
Tallahassee, Florida 32306-4390, USA}

\author{E. S. Choi, E. Jobiliong, J. S. Brooks, D. L. Eaton}

\affiliation{Department of Physics, and National High Magnetic
Field Laboratory, Florida State University, Tallahassee, Florida
32306-4390, USA}

\begin{abstract}
Resistivity measurements are reported for single crystals of
Mn$_{12}$-Acetate and Fe$_8$Br$_8$. Both materials exhibit a
semiconductor-like, thermally activated behavior over the 200-300
K range. The activation energy, $E_a$, obtained for
Mn$_{12}$-Acetate was 0.37 $\pm$ 0.05 eV, which is to be
contrasted with the value of 0.55 eV deduced from the earlier
reported absorption edge measurements and the range of 0.3-1 eV
from intramolecular density of states calculations, assuming
$2E_a$= $E_g$, the optical band gap. For Fe$_8$Br$_8$, $E_a$ was
measured as 0.73 $\pm$ 0.1 eV, and is discussed in light of the
available approximate band structure calculations. Some plausible
pathways are indicated based on the crystal structures of both
lattices. For Mn$_{12}$-Acetate, we also measured
photoconductivity in the visible range; the conductivity increased
by a factor of about eight on increasing the photon energy from
632.8 nm (red) to 488 nm (blue). X-ray irradiation increased the
resistivity, but $E_a$ was insensitive to exposure.
\end{abstract}
 \maketitle

\vspace{1in}

\renewcommand{\thesection}{\Roman{section}}
\renewcommand{\thesubsection}{\thesection.\alph{subsection}}
\section{Introduction}

The magnetic molecules
[Mn$_{12}$O$_{12}$(CH$_3$COO)$_{16}$(H$_2$O)$_4$]$\cdot$2CH$_3$COOH$\cdot$4H$_2$O,
abbreviated Mn$_{12}$-Ac \cite{TLis}, and
{[(C$_6$H$_{15}$N$_3$)$_6$Fe$_8$($\mu_3$-O)$_2$($\mu_2$-OH)$_{12}$]Br$_7$(H$_2$O)}Br$\cdot$8H$_2$O,
in short Fe$_8$Br$_8$ \cite{weig}, have been the focus of
extensive studies since it was discovered that they exhibit the
rare phenomenon of macroscopic quantum tunneling (MQT)
\cite{QTM1,QTM2,QTM3}. As has now been well established
\cite{QTM1,QTM2,QTM3,mag1,EPR5,EPR6,a2,SMM,Loss,Chud,Werns,Luis,Fern},
both of these compounds have a net total spin S = 10, and can be
grown as high quality single crystals
\cite{TLis,weig,mag1,EPR5,EPR6}. The evidence for MQT consisted of
the following observations: (a) below a certain temperature, known
as the blocking temperature, T$_B$ (2.7 K for Mn$_{12}$-Ac and 1 K
for Fe$_8$Br$_8$), their magnetic hysteresis loops exhibited sharp
steps at regular intervals (about 0.46 tesla (T) for Mn$_{12}$-Ac,
and 0.24 T for Fe$_8$Br$_8$), when the field was applied along the
easy axes \cite{QTM1,QTM2,QTM3,mag1}, and (b) the magnetization
relaxation rate became temperature independent at low temperatures
\cite{QTM1,QTM2,QTM3,mag1}. Furthermore, this quantized hysteretic
behavior was found also for very dilute samples, such as frozen
into organic solvents \cite{a2}. This observation implies that the
hysteresis loop is a property of every single molecule, rather
than that of a macroscopic domain, hence they have been described
as single molecule magnets (SMM's) \cite{SMM}. It can thus be
expected that these SMM's hold the potential for becoming an
integral part of a molecular-size memory device\cite{a2}.
Mn$_{12}$-Ac has also been proposed as a potential candidate for a
quantum computing element\cite{Loss}.

To advance our understanding of these materials for possible
applications, it is important to understand their electrical
conductivity behavior. Although these materials appear to be
insulating, they are single crystalline materials with a unique
configuration of large molecular units containing transition metal
ions and polarizable subunits, nested in a bridging network.  One
might thus expect some sort of semiconducting behavior, albeit
with high resistivity.  Interestingly, information about the
electrical conductivity is not yet available for either compound,
despite the fact that Mn$_{12}$-Ac and Fe$_8$Br$_8$ have been
studied by dielectric relaxation \cite{dielec1}, far-infrared
absorption under applied magnetic fields \cite{IR1}, Raman
scattering \cite{Ram1,Ram2,Ram3}, micro-Hall techniques
\cite{HallMn12a,HallMn12b}, micro-SQUID magnetometry
\cite{usquid1,Werns}, EPR
\cite{EPR5,EPR6,add1,add2,add3,Barra,Blinc,Park}, NMR
\cite{a26,a27,a28,a29,a30,a31,a32}, specific
heat\cite{a33,a34,a35} and magnetization measurements
\cite{QTM1,QTM2,QTM3,mag1}, neutron scattering
\cite{neut1,neut2,neut3,neut4}, and optical absorption
\cite{Opt1}. We note, however, that from optical absorption
measurements Oppenheimer \emph{et al.} \cite{Opt1} have deduced
optical excitation band gaps, $E_g$, of 1.1 and 1.75 eV for the
minority (inner tetrahedron) and majority (crown) spin systems in
Mn$_{12}$-Ac, respectively. These values were considered
comparable with corresponding theoretical estimates of 0.45 and
2.08 eV by Pederson and Khanna \cite{the10} and of 0.85 and 1.10
eV by Zeng \emph{et al.} \cite{Zeng}. In the present investigation
we have carried out electrical conductivity measurements on single
crystals of both Mn$_{12}$-Ac and Fe$_8$Br$_8$ over the
temperature range of 77 K to 300 K. Confirmatory measurements were
made using \emph{ac} dielectric techniques. The results show that
both compounds exhibit fairly clear semiconducting behavior
(200-300 K) with distinctly different transport activation
energies.  It should be noted that in an intrinsic semiconductor,
the activation energy (or gap) measured via conductivity is to be
compared with 1/2 $E_g$, and thus for Mn$_{12}$-Ac the agreement
with the optical data are satisfactory.

We also describe photoconductivity over the visible range and
X-ray damage investigations on Mn$_{12}$-Ac to further probe the
nature of electrical transport in these materials. The measured
photoconductivity exhibits a significant wavelength dependence.

\section{Experimental}

Long black rectangular crystals of Mn$_{12}$-Ac were synthesized
following the procedure of Lis \cite{TLis}.  The crystals
typically grew to dimensions of about 0.6 x 0.6 x 2.8 mm$^3$. High
quality single crystals of Fe$_8$Br$_8$ were prepared by the
method described in the literature \cite{weig}.  The Fe$_8$Br$_8$
crystals grew as dark brown orthorhombic plates of about 4.0 x 6.0
x 0.5 mm$^3$.  The samples have been routinely monitored for
quality by NMR, X-ray diffraction, and magnetization measurements.

\begin{figure}
\includegraphics[width=.6\textwidth]{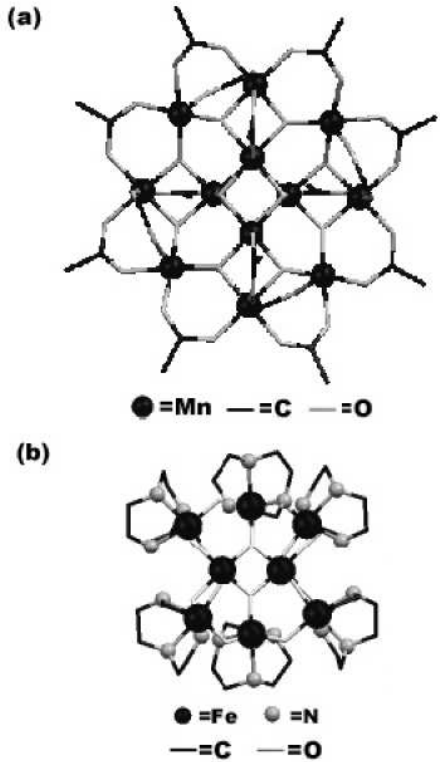} \caption{(a) Structure of Mn$_{12}$-Ac
with acetate ligand. The H$_2$O molecules are omitted for clarity
\cite{TLis}. (b) Structure of Fe\(_{8}\)Br\(_{8}\) showing the
1,4,7 triazacyclononane ligand. The Br atoms are omitted for
clarity \cite{weig}.}\label{cryst}
\end{figure}

\emph{DC} resistance measurements were conducted under either a
constant voltage or a constant current mode using a conventional
four probe technique.  A high input impedance (2$\times 10^{14}$
Ohm) electrometer was used to measure the voltage drop across the
sample when constant current was applied. Currents were typically
in the 0.1 to 10 nA range, and voltages were generally 100 V or
less. The current-voltage characteristics were periodically
checked to verify ohmic behavior. \emph{AC} conductance
measurements were made with a standard \emph{ac} impedance bridge
technique. A capacitive electrode configuration was made by
painting two flat parallel surfaces of a sample with conductive
(silver or graphite) paste. The capacitive and dissipative signals
were detected by a lock-in amplifier with an excitation frequency
of about 8 kHz. In all cases the measurements were made under
vacuum in a temperature controlled probe.

Photoconductivity measurements were made on Mn$_{12}$-Ac using a
He-Ne laser for red (632.8 nm) and Argon laser for blue (488 nm)
and green (514 nm) light. The light intensity was calibrated
before each measurement. Photocurrent was measured using a lock-in
amplifier while the sample was under \emph{direct current} bias
and illuminated by chopped light. A more detailed description of
the experiment is published elsewhere\cite{brooks}.  For the X-ray
experiments, Mn$_{12}$-Ac crystals were irradiated with 40 kV, 40
mA, Cu $K_\alpha$ radiation at room temperature in order to
observe the effects of defects.

\section{Results}
\subsection{Conductivity of Mn$_{12}$-Ac}

\begin{figure}[bp]
\epsfig{file=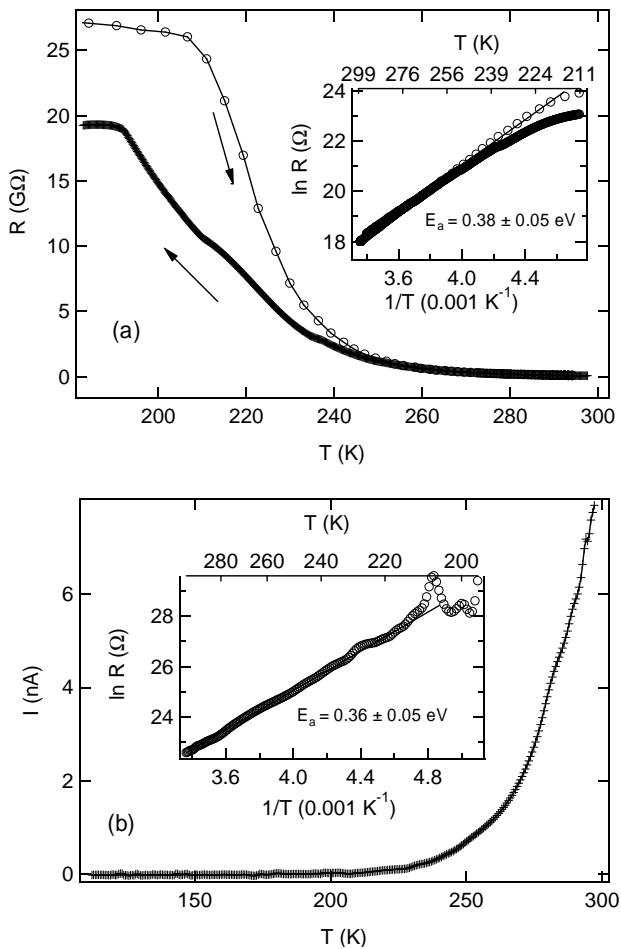, bb=79 64 518 723
clip=,width=1.0\linewidth} \caption{(a) Temperature dependence of
resistance $R(T)$ of Mn$_{12}$-Ac measured at a constant \emph{dc}
condition. The arrows indicate cooling and warming curves. Inset :
$\ln R$ vs $1/T$ curve yields $E_a$ = 0.38 $\pm$ 0.05 eV. (b)
Temperature dependence of measured current under a constant
voltage bias (50 V). Inset : $\ln R$ vs $1/T$ curve where $E_a$ =
0.36 $\pm$ 0.05 eV}\label{temp}
\end{figure}

The temperature dependence of the resistance $R(T)$ of a
Mn$_{12}$-Ac sample is shown in Fig. \ref{temp}(a) for a
4-terminal, constant current configuration. The resistivity values
are on the order of $10^9$ $\Omega$ cm at room temperature, and
increase rapidly in an activated manner upon cooling. Below about
200 K, ohmic equilibrium is lost due to the high resistance
values. Thus the Arrhenius analysis was limited to temperatures
above this temperature. Over the 200-300 K range, $\ln R$ exhibits
a linear dependence as a function of 1/T as shown in the inset,
characteristic of a semiconducting system with a well defined band
gap where $R(T) \approx \exp (E_a / k_B T)$, and $E_a$ is the
thermal activation energy. From the slopes of curves of the inset,
$E_a$ is estimated to be 0.38 $\pm$ 0.05 eV.

Fig. \ref{temp}(b) shows the temperature dependence of the current
for the constant voltage (50 V) bias condition. As the resistance
of the sample increases with decreasing temperature, the current
rapidly decreases, and is unmeasurable below $\sim$ 210 K. The
corresponding $\ln R$ vs $1/T$ curve is shown in the inset. The
linear dependence yields a value of $E_a$ = 0.36 $\pm$ 0.05 eV.

\begin{figure}[tp]
\includegraphics[width=.5\textwidth]{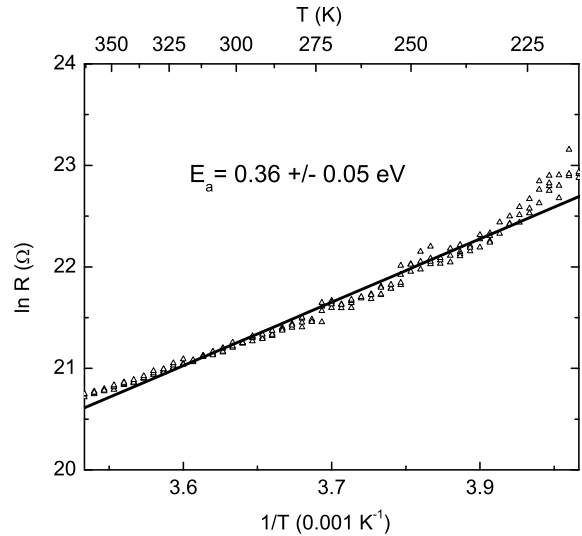}
\caption{$\ln R(T)$ vs $1/T$ curve of Mn$_{12}$-Ac measured with
the \emph{ac} impedance bridge technique. The sample was cooled
from room temperature to 200 K.  The solid line is for the curve
resulting in $E_a$ = 0.36 $\pm$ 0.05 eV. }\label{bridge}
\end{figure}

Fig. \ref{bridge} shows the $\ln R$ vs $1/T$ curve of a
Mn$_{12}$-Ac sample obtained by the impedance bridge technique.
The sample was cooled from room temperature to 200 K. The solid
line corresponds to $E_a$ = 0.36 $\pm$ 0.05 eV. The resistance
again shows activated behavior but the linear relation is not so
clear as in the \emph{dc} resistance case.

\begin{figure}
\includegraphics[width=.5\textwidth]{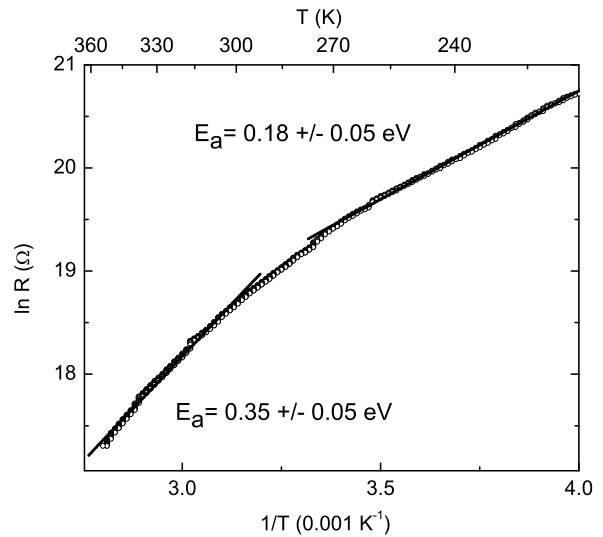}
\caption{Mn$_{12}$-Ac sample which was heated to 350 K and then
cooled down to 200 K.  Straight lines show  an $E_a$ = 0.35 $\pm$
0.05 eV for the high temperature region, and 0.18 $\pm$ 0.05 eV
for the low temperature region.}\label{heat}
\end{figure}

A significant observation was that the Mn$_{12}$-Ac crystals lose
solvent upon heating above 300 K.  One sample was heated to 350 K
and then cooled down to 200 K.  The plot of $\ln R$ vs $1/T$
yielded two separate straight lines as can be noted from Fig.
\ref{heat}. The activation energy at the higher temperature range
was 0.35 $\pm$ 0.05 eV, while the $E_a$ was 0.18 $\pm$ 0.05 eV for
the lower temperatures, after heating. Thus, care must be taken
not to heat the samples above 300 K or so.

\subsection{Photoconductivity of Mn$_{12}$-Ac}

\begin{figure}[bp]
\epsfig{file=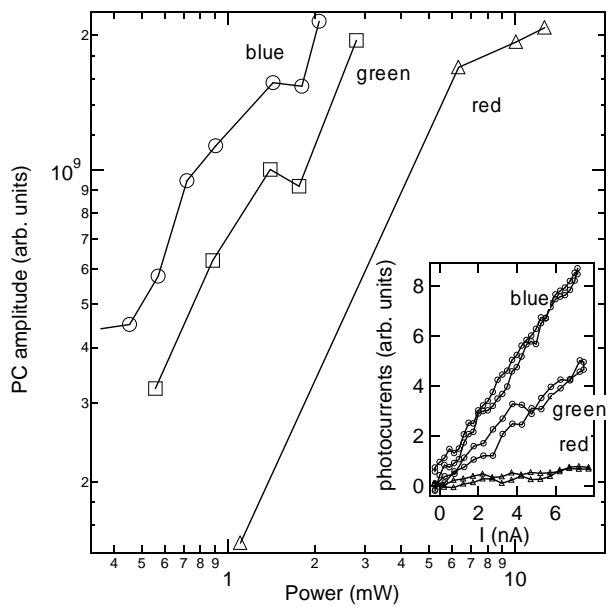, bb=63 304 505 728
clip=,width=1.0\linewidth} \caption{Photoconductivity signal of
Mn$_{12}$-Ac as a function light intensity induced by different
wavelengths of light (red: 632.8 nm, green: 514 nm and blue: 488
nm). The inset shows the \emph{ac} photocurrent as a function of
applied \emph{dc} current when the light intensity is about 1 mW.
Data for both increasing and decreasing \emph{direct current} bias
are shown.}\label{photo}
\end{figure}

Photoconductivity (PC) was measured on Mn$_{12}$-Ac using the
\emph{ac} component of the  photocurrent for chopped laser light
illumination. This was done by biasing the sample with different
values of \emph{dc} current \cite{brooks}. Fig. \ref{photo} shows
the dependence of the PC on the intensity (power) of the laser
radiation at the three wavelengths used, 632.8 nm (red), 514 nm
(green), and 488 nm (blue).  PC is seen to increase with photon
energy.  The increase is about a factor of eight when going from
632.8 nm to 488 nm.  Clearly this enhancement must relate to the
creation of charge carriers by the photons, or to the increase in
temperature due to light absorption, or both.  A simple thermal
mechanism is not supported by the earlier UV-visible absorption
data of Oppenheimer \emph{et al.} \cite{Opt1} on Mn$_{12}$-Ac. The
spectra show a gradual increase in absorption with photon energy
and the absorption edge was estimated to be about 1.1 eV. However,
over the 632.8 to 488 nm range, the absorption was nearly (within
a factor of two) constant, while the PC increases by a factor of
eight.  These considerations argue against a major role of thermal
heating in the mechanism of the observed PC enhancement.  The
effect is thus ascribed to an enhancement of charge carriers due
to optical absorption.

\subsection{Effect of X-ray irradiation on Mn$_{12}$-Ac}

Recently, Hernandez \emph{et al.} \cite{Hern} observed an increase
in the magnetization tunneling rate of Mn$_{12}$-Ac caused by
defects in the lattice as a result of X-ray irradiation and heat
treatments. In order to probe the possible role of defects in the
transport properties of Mn$_{12}$-Ac, we have also carried out an
X-ray irradiation study. As the irradiation dose was increased
from 2 hr to 20 hrs, the overall resistivity of the sample
increased, but a plot of $\ln R$ vs $1/T$ for the different
exposure times showed the activation energies remained fairly
constant.  The radiation-induced defects thus seem to act as
trapping sites for the carriers.  This effect is further discussed
in Section IV.

\subsection{Conductivity of Fe$_8$Br$_8$}

\begin{figure}
\includegraphics[width=.5\textwidth]{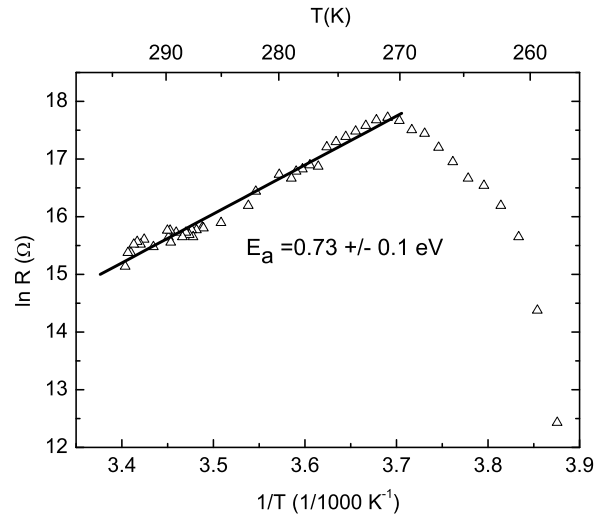}
\caption{Plot of $\ln R(T)$ vs. $1/T$ for Fe$_8$Br$_8$. The solid
line yields a value of $E_a$ $\approx$ 0.73 $\pm$ 0.1
eV.}\label{fe8res}
\end{figure}

Preliminary temperature dependent conductivity measurements have
also been carried out on single crystals of Fe$_8$Br$_8$. Figure
\ref{fe8res} shows a typical $\ln R(T)$ vs. $1/T$ plot. The slope
of the line yields a value of $E_a$ = 0.73 $\pm$ 0.1 eV, which is
seen to be significantly higher than that of Mn$_{12}$-Ac ( 0.37
$\pm$ 0.05 eV). This is discussed later in terms of the bonding of
Fe$_8$Br$_8$ and Mn$_{12}$-Ac.  At present no optical data are
available for Fe$_8$Br$_8$ for comparison with the conductivity
measurements.

Table I summarizes the conductivity results in comparison to the
optical data \cite{Opt1} and theoretical calculations
\cite{Zeng,fethe,the10}.

\section{Discussion}

\begin{figure}
\includegraphics[width=.5\textwidth]{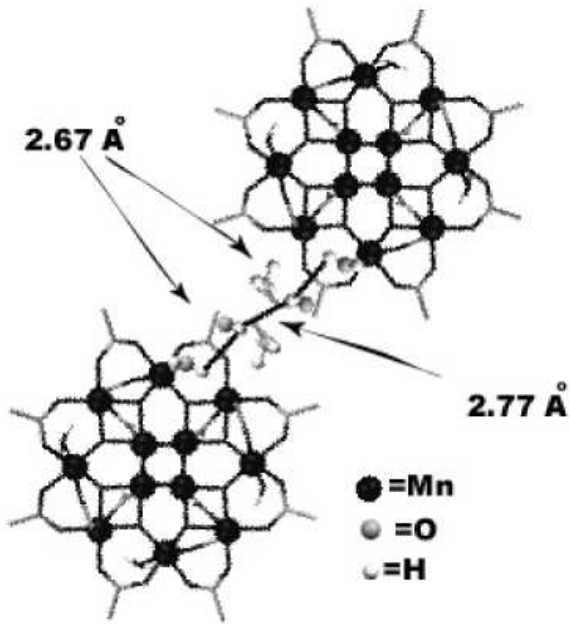}
\caption{Schematic of proposed conduction path between two
Mn$_{12}$-Ac molecules.}\label{mn12}
\end{figure}

\begin{figure}
\includegraphics[width=.5\textwidth]{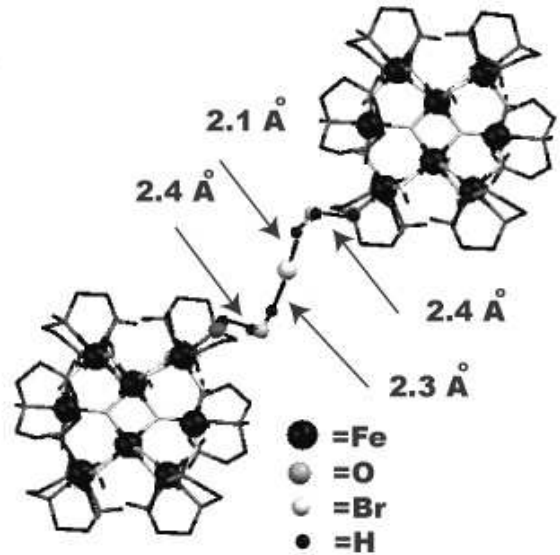}
\caption{Conduction pathway between two Fe$_8$Br$_8$ molecules.}\label{fe8}
\end{figure}

The main result of this study of the electrical transport in these
SMM-type single crystalline materials is that they exhibit
thermally activated conductivity characteristic of a gapped
semiconductor over the range of 200-300 K. Photoconductivity
measurements also support their description as gapped
semiconductors. However, the precise nature of the carrier
transport is difficult to determine, since generally measurements
over many orders of magnitude in temperature are necessary to
establish the functional dependence of $R(T)$. In order to
understand the conduction pathway, we examined the connectivity
between the neighboring Mn$_{12}$-Ac and Fe$_8$Br$_8$ clusters.
Figures \ref{mn12} and \ref{fe8} show several transport scenarios
are possible. First, as a result of the crystalline lattice, one
can expect that there will be a band structure with gaps. At
present, only the electronic structure of the clusters has been
computed\cite{the10,Zeng}.  For a band-gapped semiconductor, one
expects the resistivity to vary as $\exp(E_a/k_BT)$, where $E_a$
is related to the optical gap $E_g$ by $E_g = 2E_a$\cite{kittel}.
However, due to the complexity of the crystal structure, it is
possible that impurities and/or disorder play significant roles in
the charge transport. For example, thermally activated hopping
between impurity sites can also give similar temperature
dependence.

\begin{table}
\caption{Comparison of $E_g$ from conductivity and optical data,
and theoretical calculations.} \vspace{0.5 cm}
\begin{tabular}{|c|c|c|c|}

    \hline
    & Conductivity & Optical & Theoretical \\ \hline
    Mn$_{12}$-Ac & 0.74 $\pm$ 0.1 $eV$ \footnotemark \footnotetext{Present Work, assuming $E_g = 2E_a$}& 1.08 $eV$ \footnotemark \footnotetext{Oppenheimer \emph{et al.} \cite{Opt1}, minority spin cluster} & 0.45 $eV$ \footnotemark \footnotetext{Pederson \emph{et al.} \cite{the10}, minority spin cluster} \\ \hline
     & & 1.75 $eV$ \footnotemark \footnotetext{Oppenheimer \emph{et al.} \cite{Opt1}, majority spin cluster} & 2.08 $eV$ \footnotemark \footnotetext{Pederson \emph{et al.} \cite{the10}, majority spin cluster} \\ \hline
     & & & 0.85 $eV$ \footnotemark \footnotetext{Zeng \emph{et al.} \cite{Zeng}, minority spin cluster} \\ \hline
     & & & 1.10 $eV$ \footnotemark \footnotetext{Zeng \emph{et al.} \cite{Zeng}, majority spin cluster} \\ \hline
    Fe$_8$Br$_8$ & 1.46 $\pm$ 0.2 $eV$ \footnotemark[1] & & 0.9 $eV$ \footnotemark \footnotetext{Pederson \emph{et al.} \cite{fethe}, minority spin cluster} \\ \hline
    & & & 0.9 $eV$ \footnotemark \footnotetext{Pederson \emph{et al.} \cite{fethe}, majority spin cluster} \\ \hline

\end {tabular}
\end{table}

If conduction is through a distribution of impurity sites,
variable range hopping (VRH) should dominate. However, the
following considerations argue against a VRH behavior.
Furthermore, the resistivity ($\sim 10^9$ $\Omega$ cm) is very
high for a typical VRH conduction system\cite{mott}. Moreover,
application of the Mott formula for VHR ($R(T) \approx \exp (T_0
/T)^\gamma$, with $\gamma=1/(1+d)$, where d is the dimensionality)
yielded T$_0$ $\approx$ 3$\times 10^9$ K (for d=3). The high
sensitivity of the VRH model to defects can be tested by
introducing them artificially, by ion implantation, or in the
present case, by X-ray irradiation. Since the $T_0$ values
obtained from the Mott formula indicate an extremely small density
of impurity sites, i.e. $N(E_F)$, a small number of additional
defects should decrease $T_0$ significantly. However, our
experimental results on irradiated samples (Sec. III.c.) indicate
that $T_0$ and $E_a$ are insensitive to the creation of defects.
Hence the X-ray investigation supports the idea of intrinsic
semiconductor-like conduction in Mn$_{12}$-Ac, and by inference,
also for Fe$_8$Br$_8$.

While we have not been able to arrive at any detailed picture of
the conduction pathways, we offer the following possibilities
based on the structure and bonding characteristics of both
lattices.  The pathway for Mn$_{12}$-Ac is based upon simple
Coulombic interactions between Mn$^{3+}$ ions, and the polar
molecules which lie between two Mn$_{12}$-Ac clusters.  A water
molecule bound to a Mn$^{3+}$ on the outer crown lies 2.67
$\mathring{A}$ away from an unbound acetate ligand, which is, in
turn, 2.77 $\mathring{A}$ away from a symmetrically equivalent,
unbound acetate ligand adjacent to the closest Mn$_{12}$-Ac
cluster as seen in Figure \ref{mn12}. The Fe$_8$Br$_8$ conduction
pathway is illustrated in Figure \ref{fe8}. The proposed pathway
between two Fe$_8$Br$_8$ clusters is through an N-H bond in the
1,4,7-triazacyclononane. The conduction pathway thus extends from
the N-H to a water (2.4 $\mathring{A}$), to a Br$^-$ (2.3
$\mathring{A}$), to a water (2.1 $\mathring{A}$), and finally to a
hydrogen (2.4 $\mathring{A}$) directly connected to the
1,4,7-triazacyclononane on the adjacent Fe$_8$Br$_8$.  It should
be noted that the proposed Mn$_{12}$-Ac conduction pathway is much
more direct than that for Fe$_8$Br$_8$. This is consistent with
the higher activation energy found for Fe$_8$Br$_8$.

\section{Summary}
We have found that both Mn$_{12}$-Ac and Fe$_8$Br$_8$ exhibit a
gapped semiconductor-like behavior in their electrical transport
properties. The limited temperature range over which the
resistance was measurable, and over which the materials are
stable, restricts a knowledge of the precise functional form of
R(T). Nevertheless, complementary photoconductivity and X-ray
irradiation studies support a model where the transport is
governed by a well-defined energy gap. The $E_a$'s have been
determined to be 0.37 $\pm$ 0.05 and 0.73 $\pm$ 0.1 eV for
Mn$_{12}$-Ac and Fe$_8$Br$_8$ respectively. Assuming an intrinsic
semiconducting behavior, they lead to $E_g$ values of 0.74 $\pm$
0.10 eV for Mn$_{12}$-Ac and 1.5 $\pm$ 0.2 eV for Fe$_8$Br$_8$.
For Mn$_{12}$-Ac, the agreement is seen to be reasonably good with
the optical band gaps for minority (inner tetrahedron) spins, and
the theoretical estimates by Pederson and Khanna \cite{the10}as
well as Zeng \emph{et al.} \cite{Zeng}. Additional optical and
theoretical data are needed for Fe$_8$Br$_8$.  At present,
calculations exist only for the molecular band-gaps, but not for
the entire lattice. Hence, we can only speculate that the
inter-cluster ligand bridges may play an important role in the
conduction mechanism. Further computations on the full crystal
band structure are thus desirable.

\section{Acknowledgements } We would like to thank Dr. X. Wei for his
assistance in the optical measurements. This work is supported by
NSF-DMR 023532, DARPA, and NSF/NIRT-DMR 0103290. The National High
Magnetic Field Laboratory is supported through a cooperative
agreement between the National Science Foundation and the State of
Florida.

\end{document}